\documentclass{PoS}
\usepackage{caption}
\usepackage{lineno}
\usepackage{upgreek}

\newcommand{\spicecore}{{SPICEcore }}

\title{Design and performance of a UV-calibration device for the \spicecore hole}

\ShortTitle{UV-Logger}

\author{
The IceCube Collaboration\footnote{For collaboration list, see PoS(ICRC2019) 1177.}\\
{$^{\dagger}$ \itshape \href{http://icecube.wisc.edu/collaboration/authors/icrc19_icecube}{http://icecube.wisc.edu/collaboration/authors/icrc19\_icecube}}\\
E-mail: \email{jbrosteankaiser@icecube.wisc.edu}
}

\abstract{
The IceCube Neutrino Observatory will be upgraded in 2022/23. For this IceCube Upgrade and the planned enlarged detector IceCube-Gen2 new optical modules are under development. One of these optical modules, the Wavelength-shifting Optical Module (WOM), uses wavelength-shifting and light-guiding techniques to measure Cherenkov photons in the UV-range. To understand the potential improvements of this new module the absorption and scattering lengths of UV light in the South Pole ice need to be measured. The measurement is done utilizing an existing borehole (SPICEcore) of 1751\,m depth. The \spicecore hole was drilled for glaciological studies and filled with a transparent antifreeze liquid to remain accessible. To measure the UV optical properties a calibration device has been designed and lowered down into the hole. The device includes a UV light source and a UV-sensitive detector. UV photons scattered back are measured and from their time distribution the scattering and absorption length are calculated. We present the design of the probe and its performance during the 2018/19 measurement campaign.

\vspace{4mm}
{\bfseries Corresponding authors:}
\speaker{Jannes Brostean-Kaiser}$^{1,2}$\\
{$^{1}$ \itshape Institut f\"ur Physik, Humboldt-Universit\"at zu Berlin, D-12489 Berlin, Germany}\\
{$^{2}$ \itshape DESY, D-15738 Zeuthen, Germany}

}

\FullConference{36th International Cosmic Ray Conference -ICRC2019-\\
		July 24th - August 1st, 2019\\
		Madison, WI, U.S.A.}

\begin{document}
\section{Wavelength-shifting Optical Module}\label{Wavelength Shifting Optical Module}
 IceCube is a cubic-kilometer neutrino detector installed in the ice at the geographic South Pole \cite{Aartsen:2016nxy} between depths of 1450\,m and 2450\,m, completed in 2010. Reconstruction of the direction, energy and flavor of the neutrinos relies on the optical detection of Cherenkov radiation emitted by charged particles produced in the interactions of neutrinos in the surrounding ice or the nearby bedrock.

 In the austral summer 2022-2023 seven additional strings will be deployed for the IceCube Upgrade \cite{ex13:2019icrc}. The new strings will include several types of new optical modules developed to calibrate the optical properties of the South Pole ice and to improve the sensitivity of IceCube. A few innovative optical modules will be installed as a test platform for IceCube-Gen2, which will extend the detector of about a factor 10, with up to 125 new strings installed into the South Pole ice \cite{Gen2:2014}.

 One of these newly developed optical modules, the Wavelength-shifting Optical Module (WOM), is designed to measure Cherekenkov radiation in the UV-range \cite{WOM:2017}. The number of emitted Cherenkov photons decreases with the square of the wavelength, which results in a higher flux in the UV-range than in the visible range.
 The current prototype consists of a 90\,cm long transparent (PMMA or quartz glass) tube with 9\,cm diameter and 0.3\,cm wall thickness. The tube is coated with a wavelength-shifting paint \cite{Hebe:2014}. The paint absorbs light between 250\,nm and 400\,nm and reemits it with a peak at 420\,nm. The reemitted light is guided via total internal reflection to one of the two PMTs connected at the top and at the bottom of the tube.

 Besides the optimized sensitive range, the WOM has a much higher signal to noise ratio than the current optical modules due to the smaller PMTs and higher effective area.
 In addition, a string with only WOMs would need a smaller borehole, reducing the costs of drilling. These advantages make the WOM a very promising new optical module.

\section{Ice Properties}
To evaluate the potential improvement of this new optical module, scattering and absorption length, well known in the IceCube wavelength range, need to be measured in the UV-range. A UV calibration device (UV logger) has been built to measure these properties.

\subsection{Absorption}
The ice at the South Pole consists originally of snow that compacted over time to firn and then ice. This process is not reproducible in the laboratory, making an in-situ measurement necessary.
The light absorption in South Pole ice has been measured down to 300\,nm with AMANDA \cite{Acke:2006}. The absorption in the ice is driven by the ice itself and impurities within. In the very deep UV-range the ``Urbach tail'' \cite{Urba:1953} occurs, resulting in a very strong absorption below 200\,nm \cite{Mint:1971}. Above 500\,nm molecular absorption by pure ice occurs. Absorptivity in this regime exhibits spectral structure due to different modes of H$_2$O molecular stretching, bending and vibration, which can be excited in the ice. Between 200\,nm and 500\,nm the ice is believed to be mostly transparent, with the absorption being only due to dust particles in the ice.

\subsection{Scattering}
The effective scattering coefficient has been measured for different depths down to 337\,nm. The scattering coefficient is highly depth dependent and largest in the bubble-dominated region above 1300\,m depth. The scattering can be described with Mie scattering at a mean scattering angle of 20$^\circ$ \cite{Acke:2006}.

\section{\spicecore hole}
The deployment of the UV-Logger took place in the South Pole ice core hole (\spicecore hole). The \spicecore hole is a 126 mm diameter open borehole at about 1\,km distance from the IceCube array \cite{spice}.
The ice core was drilled out in 2\,m long parts down to a depth from 1750\,m. The hole was filled with Estisol 140. Estisol is a synthetic ester fluid that does not freeze in the South Pole environment. It has a very similar density as the South Pole ice which prevents the hole from collapsing. The hole is open for measurements of the South Pole ice. Other devices deployed in the \spicecore hole are the Luminescence Logger \cite{Anna:2019} and the Camera System \cite{Cars:2019}.

\section{Development of the a UV-Calibration device}
The UV calibration device was designed as an in-situ probe to measure the ice properties in the \spicecore hole. It consists of a light source, a UV sensitive detector and readout electronics. The light source sends out pulses at different wavelengths with nanosecond width into the ice. The detector measures the photons that are scattered back and records their arrival time with nanosecond resolution. From the time distribution one can derive the absorption and the scattering lengths. The components are contained in a quartz glass vessel \footnote{ provided by the company Nautilus Marine Service GmbH \href{https://www.nautilus-gmbh.com/de/}{website}}.
\begin{figure}
    \centering
\includegraphics[width=0.55\textwidth]{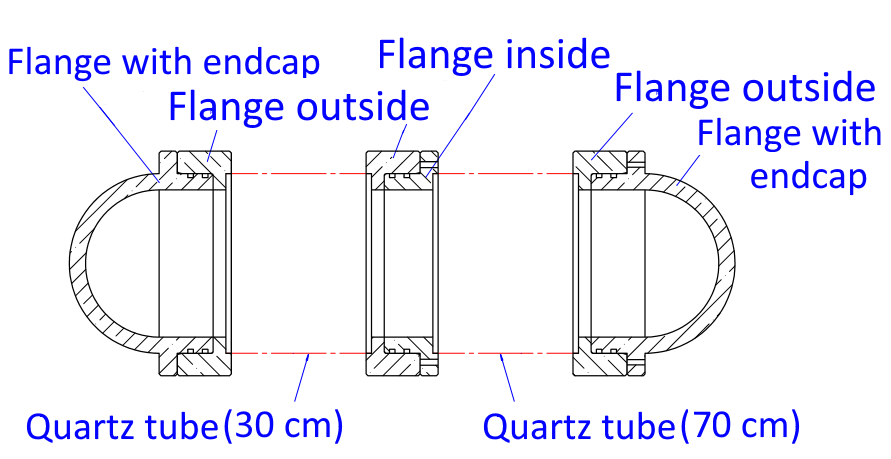} \hfill \includegraphics[width=0.4\textwidth]{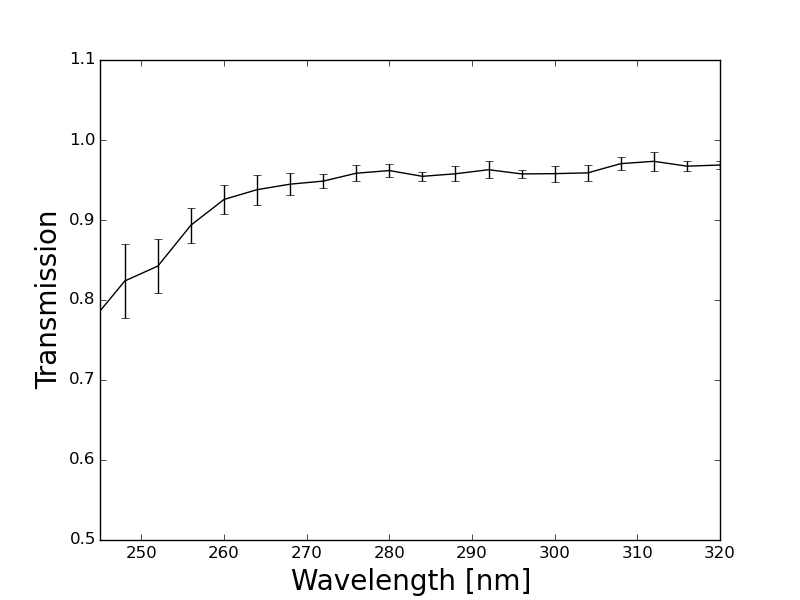}
\captionof{figure}{a) Sketch of the quartz glass housing provided by the company Nautilus$^1$. b) Transparency of the quartz glass in dependency of the wavelength. \label{house}}
\end{figure}
 Figure \ref{house} a) shows a drawing of the logger housing. It has two endcaps, one middle flange made from titanium and two quartz glass tubes. The lower glass tube, containing the electronics and the light source, is 30\,cm long. The second glass tube is 70\,cm long and houses the detection unit. Both tubes have a wall thickness of 7\,mm and are made of quartz glass because of its small natural radioactivity and its high transparency in the UV-range. The transparency for different wavelengths can be seen in Figure \ref{house} b).

 \subsection{Light source assembly}
\begin{figure}
\centering
\includegraphics[width=0.2\textwidth]{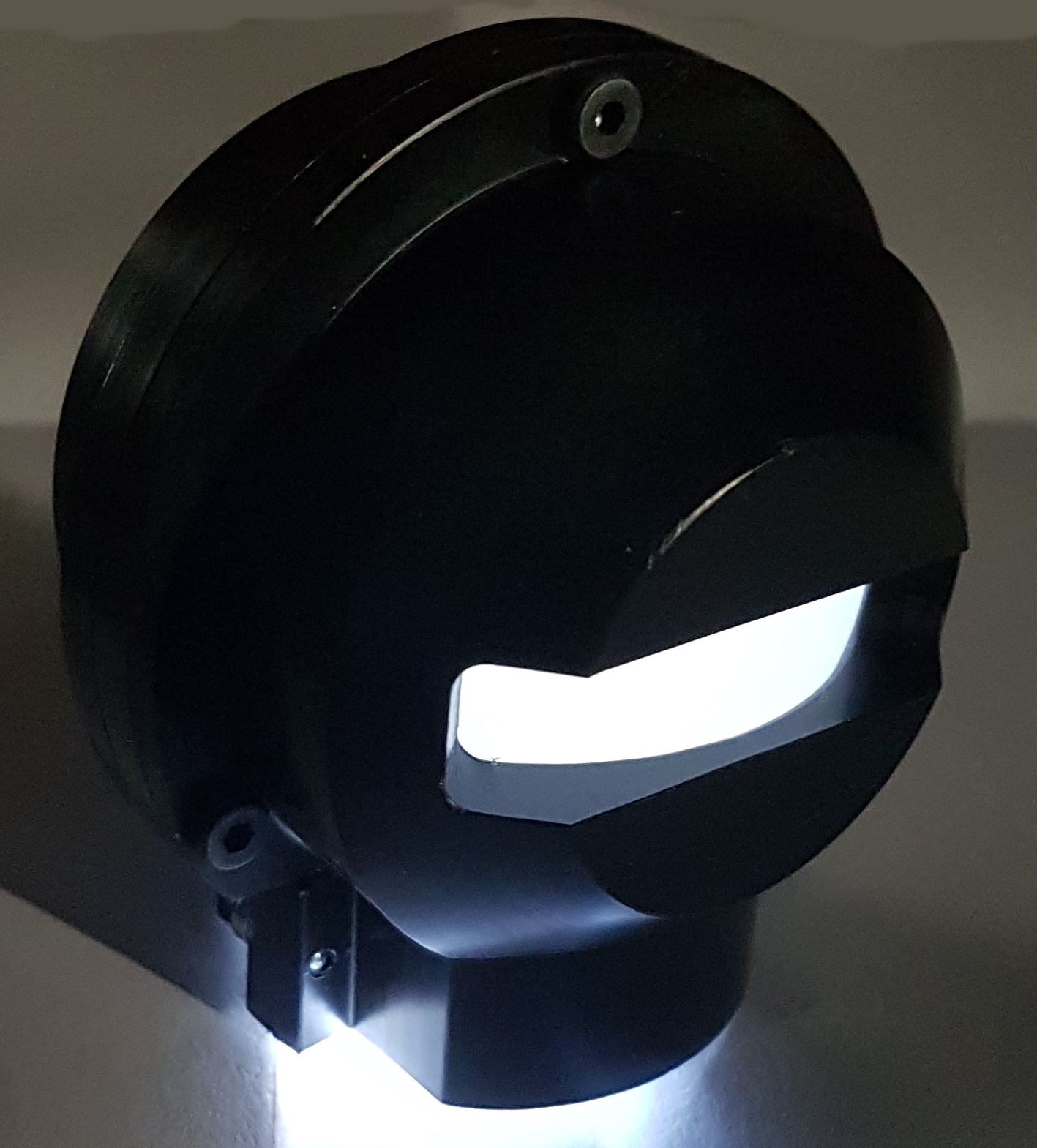} \hfill \includegraphics[width=0.79\textwidth]{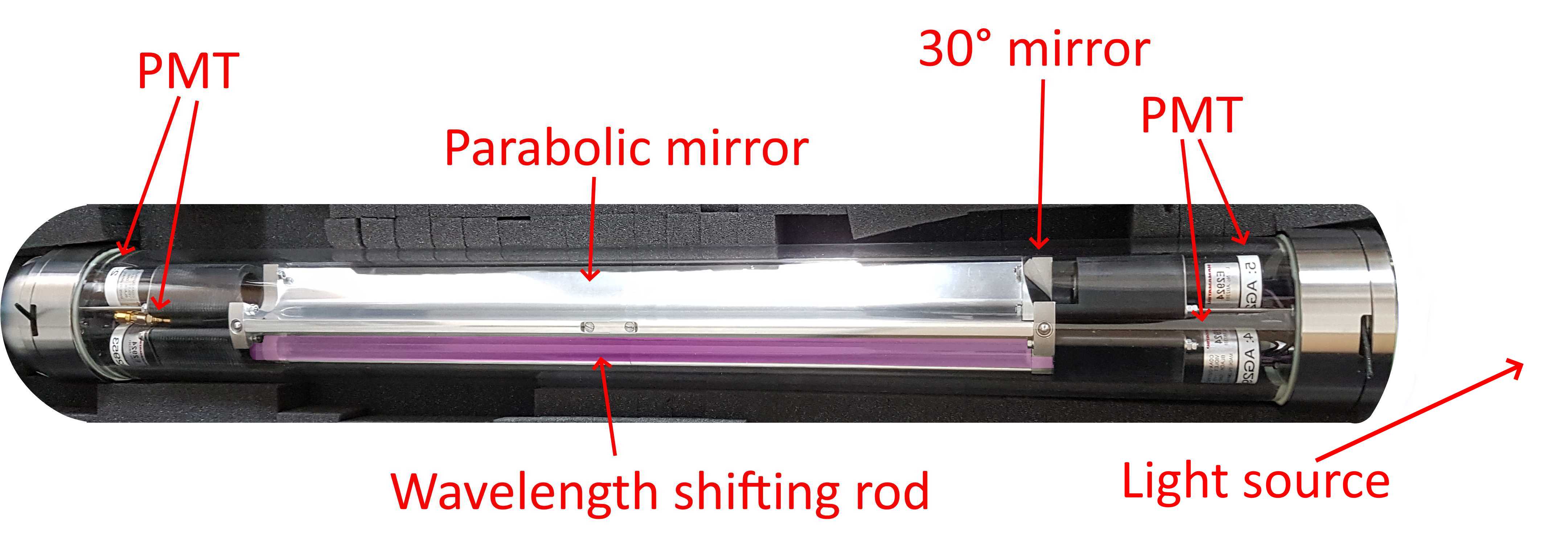}
\captionof{figure}{a) Semi-transparent integrating sphere over the LEDs. b) Detection part of the UV-Logger with the 30$^\circ$ mirror and one of the two wavelength-shifting rod visible.\label{house2}}
\end{figure}

In Figure \ref{house2} a) a picture of the integrating sphere on the flasher board is shown. The light source consists of two LEDs (278\,nm and 400\,nm) mounted on a board (flasher board) with an semi-transparent integrating sphere surrounded by a blackened housing \cite{POCA:2019}.
The housing has a slit to create a light beam with $90^\circ$ opening angle in the azimuth and $10^\circ$ opening angle in the zenith. The light beam shines into the ice perpendicular to the glass housing. In this light beam the photons have no preferred direction. 
The intensity of the LEDs is adjustable.
 \subsection{Detector assembly}
 The detector assembly is contained in the longer quartz vessel and is separated longitudinally  in three \textit{segments} by parabolic aluminum mirrors.
 Two of the segments contain a wavelength shifting rod (2\,cm diameter, 50\,cm length), using the same technology as the WOM \cite{Hebe:2014}, connected on each side to a photomultiplier tube\footnote{Hamamatsu R1924A (\href{https://www.hamamatsu.com/us/en/product/type/R1924A/index.html}{website}) operated at a gain of about $10^6$} (PMT). One of the two wavelength shifting rods is shown in figure \ref{house2} b) on the lower side of the detector. The photons entering the tubes are shifted in wavelength by the paint and guided via total reflection inside of the rods to the PMTs where they are read out. The time delay and spread due to the wavelength shifting paint is in the range of several nanoseconds.

 The third segment does not contain a rod, but also contains two PMTs to measure the photons without wavelength shifting. Instead of a rod an additional mirror (at 30$^\circ$ angle compared to the segment longitudinal axis) is positioned below the PMT nearer to the light source to deflect photons into the PMT.
 Figure \ref{house2} b) shows the mirror on the right side. In a simulation (described below) it can be seen, that most of detected photons would arrive a few centimeter below the upper PMT with 60$^\circ$ zenith angle. This simulation gives the position of the mirror. These two PMTs measure the scattered light without the wavelengthshifting and light guiding inside of the rod, but directly. The PMTs without rod can therefore measure only photons down to 300\,nm wavelength. Photons in the deeper UV range can only be detected at the PMTs with rod.

 The detection of photons without wavelength shifting provides a much better time resolution which is important to get the correct shape of the rising edge of the time distribution. From the simulations (Section \ref{Simulation}) it can be seen that the rising edge is highly correlated to the scattering length and has a width of only a few nanoseconds.
 \subsection{Readout}
 The readout system is an FPGA-based data acquisition. It can measure timestamps with a discriminator to get the time differences between the trigger of the light source and the arrival of backscattered photons.
For each light pulse only one timestamp can be recorded for each PMT. The light intensity needs to be adjusted so that on average only one photon gets scattered back into the detection unit. Photons arriving after the first photon would alter the waveform affecting the measurement of the arrival time of the first photon.
As a cross check 1\,$\upmu$s long waveforms are periodically recorded with a Domino Ring Sampler.

\section{Simulation}\label{Simulation}
\begin{figure}
     \centering
\includegraphics[width=0.4\textwidth]{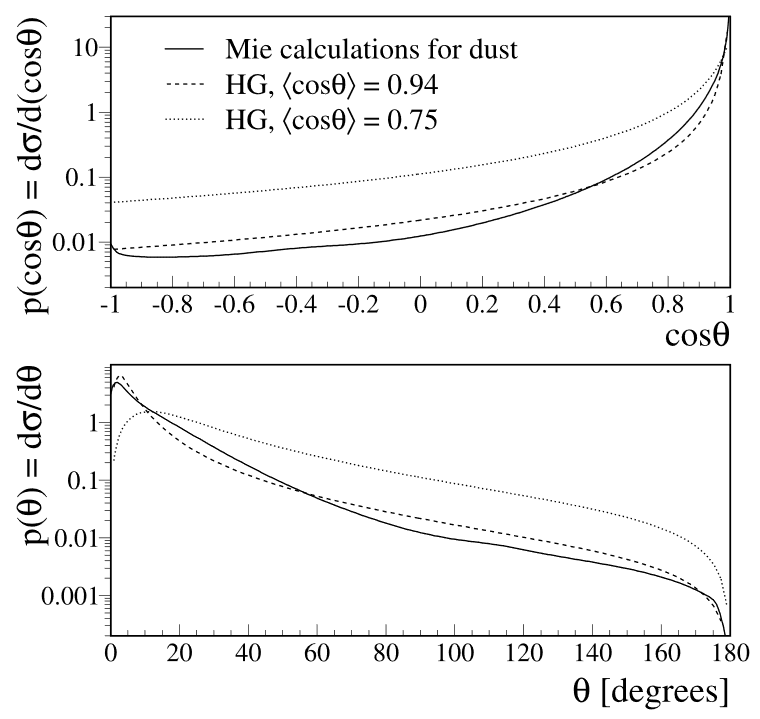} \hfill
\includegraphics[width=0.38 \textwidth]{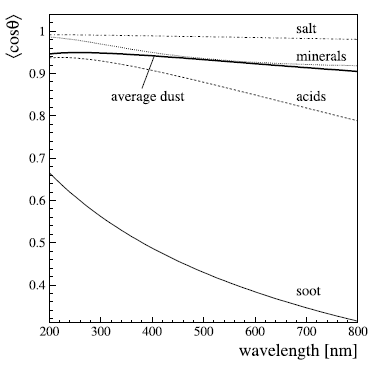}
\captionof{figure}{ a) Comparison of the Mie and the Henyey-Greenstein distributions at different mean scattering angles. b) Mean cosine of the scattering angle for different materials in dependence of the wavelength. \cite{Acke:2006} \label{mie} }
 \end{figure}

The experiment design is based on a custom developed simulation, initially focused on light at 250 nm and 400 nm wavelength. The Mie-scattering is approximated with a Henyey-Greenstein function. Figure \ref{mie} a) shows the difference between the Mie-scattering and the Henyey-Greenstein distribution for different mean angles and figure b) shows the average scattering angle for different materials. The scattering in the deep ice is assumed to be mostly dust, resulting for wavelengths between 250\,nm and 400\,nm in a mean scattering angle of $ \langle \cos \theta \rangle = 0.95 $.

To understand the requirements on the experiment and the setup simulations for light emission with a wavelength of 250\,nm and 400\,nm have been done. The detector design is simplified to an cylinder with the diameter of the hole and 50\,cm length. The transmission through Estisol and the raytracing is not implimented in the simulation. To correct for the time delay due to the wavelength shifter and the transit time of the PMT an additional randomized, depth dependent offset in the range of several nanoseconds is added to the arrival time.

\begin{figure}
    \centering
    \includegraphics[scale=0.7]{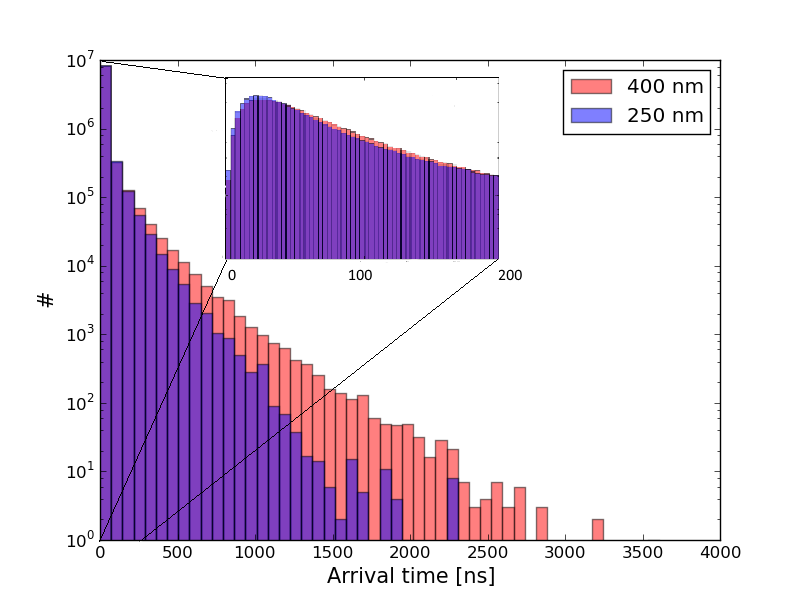}
    \captionof{figure}{Simulation for a wavelength of 250\,nm with an absorption length of $l_{\mathrm{a}250} = 50$\,m and a scattering length of $l_{\mathrm{s}250} = 50$\,cm and for a wavelength of 400\,nm with $l_{\mathrm{a}400} = 85$\,m  and $l_{\mathrm{s}400} = 77$\,cm. \label{sim} }
\end{figure}

At 400 nm absorption length and scattering length are set to $l_{\mathrm{a}400} = 85$\,m and $l_{\mathrm{s}400} = 77$\,cm.
These parameters are given in \cite{Acke:2006} along with formulas to estimate the absorption and scattering lengths for other wavelengths. With this formulas the parameters for a emitted wavelength of 250\,nm can be calculated to $l_{\mathrm{a}250} = 50$\,m and $l_{\mathrm{s}250} = 50$\,cm. The two timing distributions can be seen in Fig.\,\ref{sim}.

Further simulations have shown that the difference in the rising edge of the distributions is mostly driven by scattering
and the falling edge is mostly driven by absorption. This behavior is also described in \cite{Acke:2006}.

\section{Measurements}
In the austral summer 2018/2019 data was collected on two days, at depths of 1056\,m, 1475\,m and 1560\,m, using both the 278 nm and the 400 nm LED at each depth.  
Every 1000$^{\mathrm{th}}$ recorded timestamp a waveform was stored to cross check the time stamp measurements.

Light was only detected by three PMTs connected to the wavelength shifting rods at 1056 m depth.
This detection is the first practical application of this new developed wavelength shifting technology.
The PMTs designed to detect the photons directly were not operational. The absorptivity of the wavelength shifting paint is very low for 400 nm. Only the 278\,nm photons could be detected.

At depths of 1475\,m and 1560\,m no backscattered light was observed in the detector. The scattering at 1056\,m depth is about ten times higher than at 1475\,m and 1560\,m, thus a possible explanation is that the light source used was not bright enough.

\section{First Data Analysis} 
\begin{figure}
    \centering
    \includegraphics[width=0.5\textwidth]{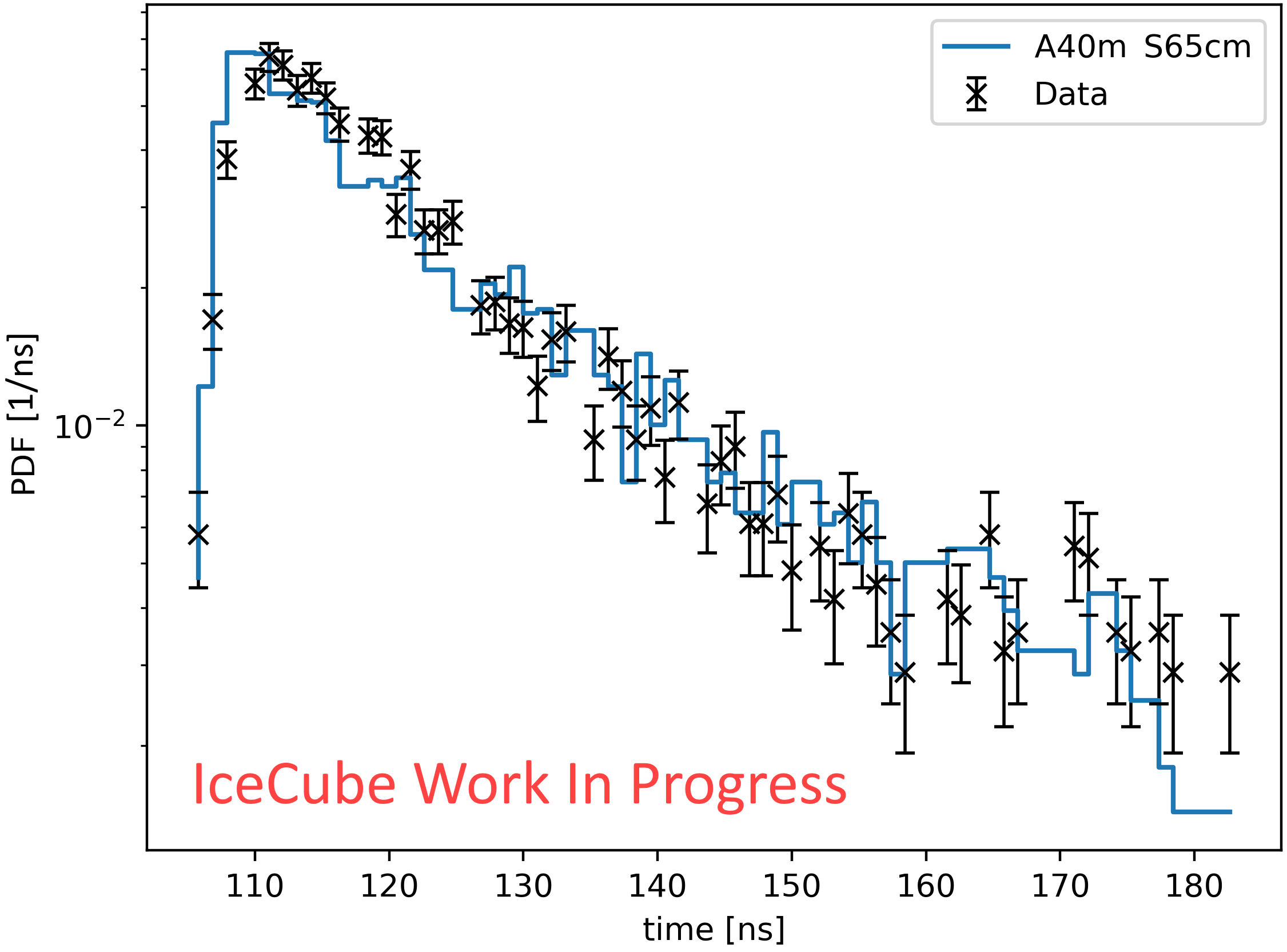}
    \caption{Probability density function of the measurement compared to a Monte Carlo Simulation for an absorption length of 60\,m and a scattering length of 75\,cm and a time offset of 92\,ns.}
    \label{meas}
\end{figure}

To analyse the measurements, the distribution of photon arrival times is compared to the simulation. The black data points in Fig.\,\ref{meas} are the the calculated PDF.
To estimate the best pair of parameters the measurement is compared to the Monte Carlo simulations from Sec.\,\ref{Simulation} with a binned maximum likelihood fit \cite{Barl:1993}. For the comparison a $\chi^2$/dof is calculated according to the formula:
\begin{equation}
    \chi^2 = \sum_{i=1}^N \frac{(d_i - a_i\cdot N_d/N_d)^2}{d_i + a_i \cdot N_d^2/N_a^2}
\end{equation}
where $N$ is total number of bins in the measurement, $d_i$ and $a_i$ are the number of events in the bin $i$ for the measurement $d$ and the Monte-Carlo simulation $a$ and $N_d$ and $N_a$ are the total number of events in the measurement and Monte-Carlo simulation. \
The time offset between the start of data acquisition and the triggering of the light source is dependent on the repetition rate and could not be measured directly so it was considered as an additional parameter and varied from 90\,ns to 95\,ns. The smallest $\chi^2$ value was found at an offset of 92\,ns. In comparison to the resolution of the measurement the offset can be assumed to be constant.

\begin{figure}
    \centering
    \includegraphics[width = 0.8\textwidth]{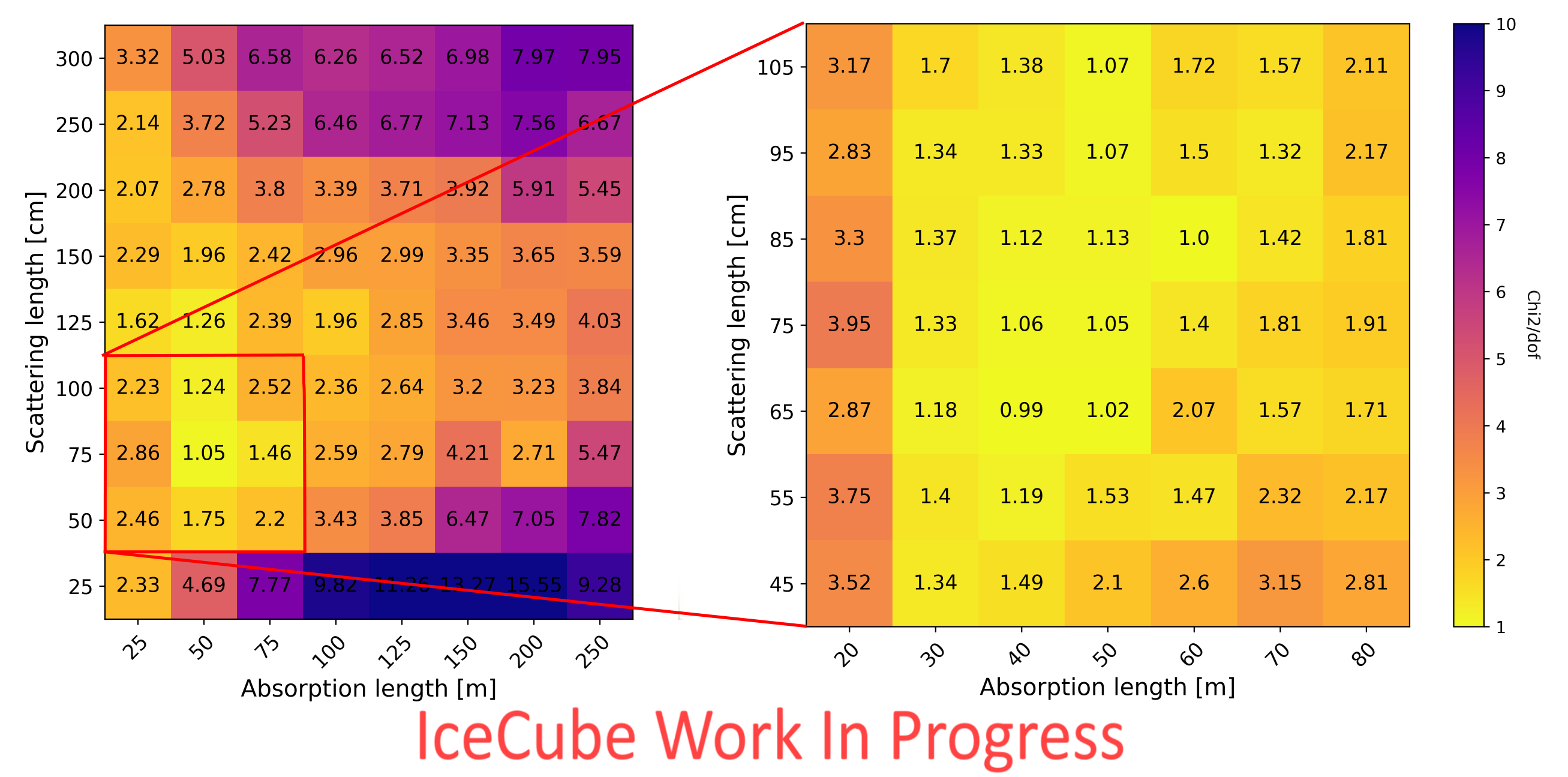}
    \caption{$\chi^2$/dof-values for different sets of parameters and an offset of 92\,ns.}
    \label{chi2}
\end{figure}

Simulations were done with scattering length ranging from 25 cm to 300 cm, absorption length ranging from 25 m to 250 m. Figure\,\ref{chi2} gives the $\chi^2$/dof-values for different sets of parameters. The offset for all calculations was set to 92\,ns. The smallest $\chi^2$-values are found at a scattering length of 65\,cm and an absorption length of 40\,m.
The main contribution of the $\chi^2$/dof are in the first 10\,ns, where the peak is located. This peak is highly dependent on the rising edge, where the time resolution of the measurement is most important.
It has to be taken in account that the scattered photons are all wavelength shifted and not detected directly.

\section{Future measurements}
The calibration device will be deployed in the 2019-2020 austral season, targeting 5 days of measurements. The logger will be improved with several new features. The 400 nm LED will be replaced by a 370 nm LED, and new LEDs will be added, so that the wavelengths available will be 250 nm, 255 nm, 278 nm, 310 nm and 370 nm.

The integrating sphere will be removed and the flasher board will be rotated 90$^\circ$, without a housing. The LEDs will shine light into the ice directly, improving the intensity significantly. With this change, light at all the wavelengths should be measurable.

The communication to the surface will be improved. For the first measurement all data was stored on a flashcard on the logger. Only the rates of each channel were transmitted to the surface.
Waveforms will be sent to the surface and will be monitored together with the rates on the channels, so that the light intensity can be adjusted to guarantee a successful measurement.

\subsubsection*{Acknowledgements}
The authors would like to thank the SPICEcore collaboration for providing the borehole, the US Ice Drilling Program, the Antarctic Support Contractor and the NSF National Science Foundation
for providing the equipment to perform the described measurement and for their support at South Pole.

\bibliographystyle{ICRC}
\bibliography{ref}

\end{document}